# Sharing Social Network Data: Differentially Private Estimation of Exponential-Family Random Graph Models

Vishesh Karwa

Pavel N. Krivitsky

Aleksandra B. Slavković

**Summary**. Motivated by a real-life problem of sharing social network data that contain sensitive personal information, we propose a novel approach to release and analyze synthetic graphs in order to protect privacy of individual relationships captured by the social network while maintaining the validity of statistical results. A case study using a version of the Enron e-mail corpus dataset demonstrates the application and usefulness of the proposed techniques in solving the challenging problem of maintaining privacy *and* supporting open access to network data to ensure reproducibility of existing studies and discovering new scientific insights that can be obtained by analyzing such data. We use a simple yet effective randomized response mechanism to generate synthetic networks under $\epsilon$-edge differential privacy, and then use likelihood based inference for missing data and Markov chain Monte Carlo techniques to fit exponential-family random graph models to the generated synthetic networks.

*Keywords*: Enron e-mail corpus, ERGM, differential privacy, missing data, randomized response, synthetic graphs

## 1. Introduction

Networks are a natural way to summarize and model relationship information among entities such as individuals or organizations. Entities are represented as nodes, the relation between them as edges and the attributes of the entities as covariates. Such a network representation has become a prominent source of scientific inquiry for researchers in economics, epidemiology, sociology and many other disciplines. However, network data very often contain sensitive relational information (e.g., sexual relationships, email exchanges, financial transactions), while the covariate information can, in some cases, be assumed to be safe to release. The social benefits of analyzing such data are significant, but any privacy breach of the relational information can cause public shame and even economic harm to the individuals and organizations involved. With the increase in the quantity of data being collected and stored, such privacy risks are bound to increase.

In tension with privacy issues is the need to allow open access to data to ensure reproducibility of existing studies and to discover new scientific insights that can be obtained by reanalyzing such data. As a concrete example of this tension, consider the famous National Longitudinal Study of Adolescent to Adult Health (Add Health) (Harris et al., 2003). Most of the data collected on individuals (nodes) are available to researchers subject to some confidentiality constraints and security requirements, e.g., as the *Restricted-Use* data (see Add Health (2009a)). Also collected were *Romantic Pairs* (relational) data (see Add Health (2009b)), analyzed by Bearman et al. (2004), for example. The constraints and the requirements on the relational data are far more severe: the former is shared on a project-by-project basis, with review and renewal every three years, and can be held on a networked server, while the latter is shared with only one researcher at a time, subject to review every year, and must be held on a computer system physically isolated from any computing networks. In other words, individual node level data – even Restricted-Use – is far easier to obtain and analyze than relational data.

In this paper, we consider the problem of limiting disclosure risk of relational information while allowing for statistical inference on networks in the context of three real-world network datasets, with primary focus on Enron e-mail exchanges network. We propose a method to release differentially private synthetic networks and evaluate the utility of fitting exponential random graph models using a missing data likelihood method.

Over the past decade, the Enron e-mail corpus (Klimt and Yang, 2004), comprising the e-mail correspondence among 158 employees of Enron Corporation between 1998 and 2002, has become a classic dataset in the area of text mining and analysis, and social network analysis. A big reason for its popularity is its uniqueness: no other lawfully obtained network data on corporate communications of this completeness and scale is available to researchers without severe restrictions. This is because such communications are often considered highly sensitive – even more so than individual-level attributes such as gender, department, and position, which are often public information or nearly so from corporate web sites, disclosures to regulators, the employees' own online public profiles (e.g., Facebook or LinkedIn), or court filings when cases like Enron's do occur. Who works for a company and in what official capacity is often much less sensitive than their communications, particularly the content, but also the "metadata" of who communicated with whom and how often. Enron data release comes from an era when privacy implications of such disclosure were only beginning to be appreciated, and it is likely that if a similar scandal were to take place today, the participants would likewise be publicly identified, but the correspondence would not be publicly disclosed.

We therefore use Enron network as our primary case study of a network dataset whose actor-level information would, in the ordinary course of things, be fairly public, but whose patterns of communications would be sensitive and therefore subject to controlled disclosure. In addition, we study two publicly available datasets and report on those in the online supplement (Karwa et al., 2016): A teenage friendship and substance use network formed from the data collected in the Teenage Friends and Lifestyle Study (Michell and Amos, 1997; Pearson and Michell, 2000) for a cohort of students in a school in Scotland, and a network formed from the collaborative working relations between partners in a New England law firm (Lazega, 2001).

## 2. Contributions of this study in relation to previous work

Limiting the disclosure risk while allowing for the data to remain useful has been the subject of many studies in statistics and data mining, and numerous techniques have been developed in the fields of statistical disclosure limitation (SDL) and privacy-preserving data mining, albeit with a limited focus on network data. For a survey on SDL methods which focus on statistical methodology to address this inherent trade-off, see for example, Fienberg and Slavković (2010) and Hundepool et al. (2012). A drawback of these techniques is that in most cases they do not offer any formal privacy guarantees – whether or not a disclosure technique makes the data "safe" to release is left to the subjective decision of data curator and the risk is highly dependent on the assumed knowledge of what additional information the potential intruder may have. Due to this, "naive" privacy-preserving methods, such as anonymization (removing the basic identifiers such as name, social security number, etc.) have been shown to fail and can lead to disclosure of individual relationships or characteristics associated with the released network (e.g., see Narayanan and Shmatikov (2009); Backstrom et al. (2007)). To overcome this risk, one needs a principled and formal way to reason about how to *measure* and *limit* the privacy risks of data release mechanisms over multiple data releases.

The framework of *differential privacy* (DP) (Dwork et al., 2006) has emerged from the theoretical computer science community as a principled way to provably limit a *worst case* disclosure risk in presence of any arbitrary external information. While disclosure risk has long been a subject of study and quantified in SDL, the DP risk is the first one that composes: the cumulative risk can be controlled over multiple data releases and it allows for a modular design of data release mechanisms. A significant amount of work on DP has been undertaken in theoretical computer science, and some in statistics, showing that any data release method that satisfies DP comes with strong worst-case privacy guarantees. We use it to meet the goal of sharing social network data, in the form of synthetic networks, while protecting the privacy of individual relationships. *Edge Differential Privacy* (EDP), in particular, considers the worst-case risk of the state of a relationship between any two individuals in the network being exposed. However, a common criticism of DP is that it may be too strong of a guarantee for statistical applications and more importantly, the primary focus of DP-based techniques is on releasing summary statistics of the data, as opposed to performing statistical inference.

To address the utility issue, we adopt ideas and techniques from missing data methods to ensure that one can perform *valid statistical inference on differentially private synthetic networks*. We focus on Exponential-Family Random Graph Models (ERGMs) (Hunter et al., 2008), because they have become the essential tool for analyzing social network data (Goodreau et al., 2009; Robins et al., 2007; Goldenberg et al., 2010). The current DP methods for network data are primarily focused on releasing noisy sufficient statistics of ERGMs, but fall short of demonstrating how to perform valid statistical inference using the noisy statistics. For example, Hay et al. (2009) propose an algorithm for releasing the degree distribution of a graph using the Laplace noise-addition mechanism along with post-processing techniques to reduce the $L_2$ error between the true and the released degree distribution. Karwa et al. (2011) release subgraph counts such as number of $k$-triangles and $k$-stars by adding noise using the smooth sensitivity framework of Nissim et al. (2007). Parameter estimation using such noisy sufficient statistics is a non-trivial task, as discussed and demonstrated in the context of a class of ERGMs known as the $\beta$-model by Karwa and Slavković (2012, 2015), and by Fienberg et al. (2010) in the context of non-existence of maximum likelihood estimators (MLEs) of log-linear models of contingency tables.

Ignoring the noise addition process, which is often done in the case of private release of summary statistics or synthetic data, can lead to inconsistent and biased estimates – as already well established in the statistics literature on the measurement error models, e.g., see Carroll et al. (2012). Motivated by the latter, Karwa and Slavković (2015) take the noise addition process into consideration and construct a differentially private asymptotically normal and consistent estimator of the $\beta$-model to achieve valid inference. However, the main technique that relies on projecting the noisy sufficient statistics onto the lattice points of the marginal polytope corresponding to the $\beta$-model does not scale well to more general ERGMs. Lu and Miklau (2014) propose to release perturbed ERGM sufficient statistics for the model of interest and propose a Bayesian exchange algorithm for recovering parameters from it. Karwa et al. (2014) were first to develop techniques for fitting and estimating a wide class of ERGMs in a differentially private manner by considering the original private network as missing, and taking a likelihood-based approach to ERGM inference from data released by privacy-preserving mechanisms.

In this paper, we expand on the work of Karwa et al. (2014), by improving both the methodology and the results, to address the above-described problem of limiting disclosure risk of relational information while allowing for statistical inference in the context of three real-world network datasets. We assume that the covariate information of the nodes is public, while the relational information is sensitive and requires protection. Our goal is to release synthetic versions of the networks ensuring strong privacy protection of the relational information while *any* statistical analyses can be performed on the synthetic datasets without sacrificing utility.

We use the framework of ERGMs for measuring utility and EDP framework for measuring disclosure risk. Directly applying EDP to real-world data exposes its limitations, and we propose to address them by varying privacy risks for potential relations (*dyads*) depending on the attributes of the nodes they connect. Finally, but crucially, we use missing data methods to perform valid inference based on these synthetic networks, allowing users to fit *any* ERGM to the disclosed data and quantify uncertainty in parameter estimates, including that introduced by the privacy mechanism. We combine ideas and methods from the computer sciences and the statistics to simultaneously offer rigorous privacy guarantees and analytic validity. More specifically, the following are the novel contributions of this paper:

(a) Motivated by the lack of utility in analyses of the *Teenage Friendship* data in Karwa et al. (2014), in Section 3 we present a generalized randomized response mechanism to release synthetic networks under $\epsilon$-edge differential privacy.
The new mechanism can handle directed graphs and allows for different levels of privacy risk for different types of dyads depending on the potential sensitivity of the connections, based on the nodal attributes.

(b) The Randomized Response mechanism for sharing network data is thoroughly analyzed both theoretically and in the case studies, specifically from an applied point of view. In Section 4, Lemma 2, we analyze the optimal parameters of the generalized randomized response mechanism introduced in the current paper. This analysis brings forth a very important limitation – Measuring disclosure risk by worst case (as in EDP) is oblivious to any asymmetry that one may wish to assign in the privacy risks. In particular, EDP does not recognize asymmetric disclosure risks to edges and non-edges or different types of edges (e.g., edges between the same gender vs different genders in a sexual network).

(c) We present an alternate privacy-preserving method that aims at overcoming this limitation of differential privacy and allows for different disclosure risks for different types of dyads. We use the *Enron* data as a case study of this new scheme to show that it performs better in terms of utility.

(d) In Section 5, we present improved MCMC algorithm used in Karwa et al. (2014). The new MCMC algorithm is based on the two-MCMC approach of Handcock et al. (2010) and is modified to handle the generalized randomized response mechanism.

The rest of the paper is organized as follows. In Section 3, we introduce differential privacy and the randomized response mechanism used to release the networks. In Section 4 we study the risk-utility tradeoff. In Section 5, we develop MCMC based likelihood inference procedures to analyze networks released by the differentially private mechanism. In Section 6, we present the Enron case study; additional case studies are presented in the supplementary material (Karwa et al., 2016). These case studies demonstrate the application and usefulness of the proposed techniques in solving the challenging problem of maintaining privacy and supporting open access to network data to ensure reproducibility of existing studies and discovering new scientific insights that can be obtained by analyzing such data. In Section 7 we discuss overall ramifications of data sharing under privacy constraints and some future directions.

## 3. Differential privacy for networks and Randomized response

In this section we set up the notation and propose a generalized randomized response mechanism with *edge differential privacy* (EDP), which measures the worst case risk of identifying any relationship when data are released in the form of a synthetic network.

Let $X$ be a random graph with $n$ nodes and $m$ edges, represented by its adjacency matrix. The adjacency matrix is a binary $n \times n$ matrix with zeros on its diagonal, such that $x_{ij} = 1$ if there is an edge from node $i$ to node $j$, 0 if there is no edge, or *non-edge* between nodes $i$ and $j$. We focus on graphs with no self-loops or multiplicitous edges, and our discussion applies equally to directed and undirected, as well as unipartite and bipartite (affiliation) graphs. Let $\mathcal{X}$ denote the set of all possible graphs of interest on $n$ nodes. The distance between two graphs $X$ and $X'$, is defined as the number of edges on which the graphs differ and is denoted by $\Delta(X, X')$.

Each node can have a set of $p$ attributes associated with it. These attributes can be collected in the form of a $n \times p$ matrix of covariates $Z$. We assume that the matrix $Z$ is known and public or has been released using an independent data release mechanism.

We are interested in protecting the relationship information in the network $X$, so we randomize the response to each dyad (potential tie) of the adjacency matrix of $X$.

### 3.1. Interactive Data Access versus Releasing Synthetic Networks

Differential privacy (DP) framework (Dwork et al., 2006) is designed to capture the worst-case risk of releasing sensitive data, and is defined with an eye towards interactive data access with focus on releasing summary statistics. The data $x$ (e.g., an observed network) is stored with a curator and the analyst requests summary statistics $\boldsymbol{g}(x)$ and receives noisy answers. Such process is repeated – each time the user requires access to the data, she has to interact with the curator. This is an *output* perturbation type algorithm which works by adding noise calibrated to the *sensitivity* of the sufficient statistic, which is a measure of change in $\boldsymbol{g}(x)$ over neighboring networks. The goal is to *mask* large changes in $\boldsymbol{g}(x)$ as $x$ changes over neighboring networks. In an interactive setting, the loss in privacy accumulates over time and the amount of noise added increases.

Non-interactive access provides an alternative approach to data sharing. In this setting, for example, by perturbing $x$ directly, the data curator may release one or more *synthetic datasets* (e.g., synthetic networks). This is an example of *input perturbation* algorithm. While in both cases of input and output perturbation, the perturbing mechanism is known publicly, one advantage of having access to synthetic dataset(s) is the support for more varied data analyses, typically greater than those only relying on the few sufficient statistics, that can be carried out by the analyst using the synthetic dataset(s) without interacting with the curator. On the other hand the dimension of $\boldsymbol{g}(x)$ is usually much smaller than that of $x$, which may mean that to achieve the same level of disclosure, each element of $x$ requires more noise than each element of $\boldsymbol{g}(x)$.

Laplace mechanism (Dwork et al., 2006) is a basic DP output perturbation mechanism for releasing any summary statistic $\boldsymbol{g}(x)$. It works by adding Laplace noise to $\boldsymbol{g}(x)$ proportional to its *global sensitivity*, which is the maximum change in $g$ over neighboring networks. Let $\boldsymbol{g}(x)$ be the number of edges in the network; the global sensitivity of $\boldsymbol{g}(x)$ is 1, since adding or removing a single edge changes the edge count by 1. For a non-trivial example, let $g(x)$ count the number of triangles. The global sensitivity in this case is $O(n^2)$ and thus very large. As an alternative mechanism, one can also add noise proportional to the so called smooth version of local sensitivity (Karwa et al., 2011).

Output perturbation mechanisms that release noisy summary statistics are not suitable for releasing synthetic graphs for estimating a large class of ERGMs for three major reasons. First, the set of sufficient statistics released by the curator defines the space of models that can be estimated. Thus, the models (and substantive questions) not anticipated by the curator cannot be fitted. Second, the noisy summary statistics are typically no longer sufficient (ancillary statistics can now provide some information about the network) and typically not usable for estimating model parameters and performing statistical inference, e.g., see Fienberg et al. (2010); Karwa and Slavković (2012). Third, the curator needs to design mechanisms for sufficient statistics (including estimating their sensitivity) on a case by case basis, which puts a considerable and possibly an insurmountable burden on the curator: calculating the smooth sensitivity of many network summary statistics is NP hard (Karwa et al., 2011). To avoid these issues, we propose using an input perturbation mechanism to release synthetic networks that satisfy DP.

Randomized response is one of the simplest examples of an input perturbation that would allow for release of synthetic data, where the input data $x$ are perturbed by a known probability distribution. A more commonly used method for releasing synthetic data is for the curator to fit a model to the data and release samples from the fitted model; there is an extensive literature on this topic, e.g., Raghunathan et al. (2003), Reiter (2003), Kinney and Reiter (2010), Slavković and Lee (2010), Drechsler (2011), Raab et al. (2016). Because the synthetic data only embodies structure in the curator's model, this, once again, requires the curator to anticipate all possible models the user of the data might want to fit. Performing model selection to choose a good model, estimating its parameters and releasing synthetic data under the additional requirements of DP largely remains an open problem, especially in the context of network data.

We propose a randomized response scheme for perturbing the edges and non-edges of the network to generate a collection of synthetic edges, without relying on a model, while satisfying DP to control the risk. Randomized response originated in survey methodology and has been used extensively to obtain answers to sensitive questions (Chaudhuri, 1987). Randomized response has also been used for statistical disclosure control when releasing data in the form of contingency tables (Hout and van der Heijden, 2002), and, in fact, in the context of contingency tables, it has been shown that randomized response satisfies a much stronger notion of privacy called Local Differential Privacy (Duchi et al., 2013).

### 3.2. Randomized Response for networks with Edge Differential Privacy

Edge differential privacy (EDP) is defined to measure the worst case disclosure risk of identifying any relationship (represented by edges) between entities (represented by nodes). Consider that any privacy-preserving mechanism can be modeled as a family of conditional probability distributions, which we denote by $P_\gamma(Y = y|X = x)$. Here, $x$ is the network that requires privacy protection, $Y$ is the random synthetic network obtained by sampling from this distribution, and $\gamma$ is a (vector) parameter of the privacy mechanism controlling the generation of $Y$ from $x$, which we assume is fixed and known.

Let $x$ and $x'$ be any two neighboring networks (i.e., they differ by one edge). EDP bounds the worst case ratio of the likelihoods of $Y$ conditional on $x$ and $x'$. More precisely, the mechanism $P_\gamma(Y = y|X = x)$ is $\epsilon$-edge-differentially private if, and only if,

$$\max_{y} \max_{x,x':\Delta(x,x')=1} \log \frac{P_\gamma(Y = y|X = x)}{P_\gamma(Y = y|X = x')} \leq \epsilon.$$

EDP requires that the distribution of data release mechanism on two neighboring networks should be close to each other. The parameter $\epsilon$ controls the amount of information leakage and measures the disclosure risk; smaller values of $\epsilon$ lead to lower information leakage and hence provide stronger privacy protection. One can show that even if an adversary knows all but one edge in the network, DP ensures that the adversary cannot accurately test the existence of the unknown edge. Wasserman and Zhou (2010) formalize this property using the notion of a hypothesis test and their result implies that there exist no hypothesis test that has any power to detect the presence (or absence) of any unknown edge, even if the adversary knows all the other edges. Another key property of DP is that any function of a differentially private algorithm is also differentially private without any loss in the disclosure risk, as measured by $\epsilon$ (Dwork et al., 2006; Nissim et al., 2007), a result we reproduce below as Lemma 1. This allows us to perform any kind of post-processing on the output of a differentially private mechanism without compromising the privacy and is a key requirement in the success of our methods.

LEMMA 1 (POST-PROCESSING DWORK ET AL. (2006); NISSIM ET AL. (2007)). *Let $f$ be an output of an $\epsilon$ differentially private algorithm applied to a graph $X$ and $g$ be any function whose domain is the range of $f$. Then $g(f(X))$ is also $\epsilon$-differentially private.*

Consider a graph with a collection of labeled nodes and dyads that represent the ties between each nodes. We apply randomized response to each dyad of the adjacency matrix of $X$. More specifically, for each dyad $(i,j)$ let $p_{ij}$ be the probability that the mechanism retains an edge if present, and $q_{ij}$ be the probability that the mechanism retains a non-edge. Algorithm 1 shows how to release a random graph $Y$ from $X$ that is $\epsilon$-edge differentially private. Note that for an undirected graph, we need to release $n(n-1)/2$ binary dyads and for a directed graph, $n(n-1)$.

---
**Algorithm 1** Dyadwise randomized response.

1: Let $x = \{x_{ij}\}$ be the adjacency matrix of $X$
2: **for** each dyad $x_{ij}$ **do**
3:     **if** $x_{ij} = 1$ **then**
4:       Let $y_{ij} = \begin{cases} 1 & \text{with probability } p_{ij} \\ 0 & \text{otherwise} \end{cases}$
5:     **else**
6:       Let $y_{ij} = \begin{cases} 1 & \text{with probability } 1 - q_{ij} \\ 0 & \text{otherwise} \end{cases}$
7:     **end if**
8:     Let $Y_{i,j} = \{y_{ij}\}$.
9: **end for**
10: **return** $Y$

---

We assume that the parameters of Algorithm 1 are public, i.e., the matrix of values of $p_{ij}$ and $q_{ij}$'s are known, otherwise the parameters of any model to be estimated from the released network will not be identifiable. This does not increase the privacy risks as the privacy protection comes from the randomness inherent in the mechanism and not in the secrecy of the parameters of the mechanism. The privacy risk of each dyad is measured by $\epsilon_{ij}$ and the worst case risk over all dyads is $\epsilon$. Proposition 1 shows that Algorithm 1 is $\epsilon$-differentially private.

PROPOSITION 1. *Let the privacy risk of each dyad $i,j$ be*

$$\epsilon_{ij} = \log \max \left\{ \frac{q_{ij}}{1-p_{ij}}, \frac{1-p_{ij}}{q_{ij}}, \frac{1-q_{ij}}{p_{ij}}, \frac{p_{ij}}{1-q_{ij}} \right\}.$$

*Algorithm 1 is $\epsilon$-edge differentially private with $\epsilon = \max_{ij} \epsilon_{ij}$.*

*Proof*  Consider two networks $x$ and $x'$ that differ by one edge, say $kl$. Let $Y$ be the output of Algorithm 1.

$$\frac{P_\gamma(Y=y|X=x)}{P_\gamma(Y=y|X=x')} = \frac{\prod_{ij} P(Y_{ij}|X_{ij})}{\prod_{ij} P(Y_{ij}|X'_{ij})} = \frac{P(y_{kl}|x_{kl})}{P(y_{kl}|x'_{kl})} = \frac{P(y_{kl}|x_{kl})}{P(y_{kl}|1-x_{kl})}$$

Note that $P(y_{kl}|x_{kl}=1) = p_{ij}^{y_{kl}}(1-p_{ij})^{1-y_{kl}}$ and $P(y_{kl}|x_{kl}=0) = (1-q_{ij})^{y_{kl}} q_{ij}^{1-y_{kl}}$. The only possible values of $y_{kl}$ are 0 or 1. Thus with some algebra, the max over all outputs is obtained by $\max \left\{ \frac{q_{ij}}{1-p_{ij}}, \frac{1-p_{ij}}{q_{ij}}, \frac{1-q_{ij}}{p_{ij}}, \frac{p_{ij}}{1-q_{ij}} \right\}$ which completes the proof.  □

For any dyad $(i,j)$, if $p_{ij}$ or $q_{ij}$ is equal to 1 or 0 we obtain $\epsilon = \infty$, which in the EDP model represents infinite risk (i.e., no privacy). Hence, to obtain finite privacy risks, no dyad can be left unperturbed: every dyad must have a positive probability of being perturbed. On the other hand if for all dyads, $p_{ij} = q_{ij} = 0.5$, then $\epsilon = 0$. This setting of parameters has 0 risk and provides the maximum possible privacy protection, but it also has 0 utility, as all the information in the original network is lost and there is no identifiability. We obtain a range of $\epsilon$ from 0 to $\infty$ for intermediate values of $p_{ij}$ and $q_{ij}$.

## 4. The Risk–Utility Trade-off

### 4.1. Optimal Randomized Response parameters and a limitation of the worst-case risk measure

Recall, the privacy risk of each dyad is measured by $\epsilon_{ij}$ and the worst-case risk is measured by $\epsilon$. Larger values of $\epsilon_{ij}$ ($\epsilon$) correspond to higher privacy risk for each dyad (higher worst-case risk). In the randomized response mechanism, there are infinitely many values of $p_{ij}$ and $q_{ij}$ that are equivalent to a fixed risk $\epsilon_{ij}$. Thus, for a fixed value of $\epsilon_{ij}$, what are the optimal values of $p_{ij}$ and $q_{ij}$? That is, for a fixed value of risk, is there a pair of $(p_{ij}, q_{ij})$ that is *better* for utility? The answer depends on how we measure utility. We want to ensure that each released dyad $Y_{ij}$ be close to $X_{ij}$ with high probability. This is equivalent to requiring $p_{ij}$ and $q_{ij}$ to be as close to 1 as possible.

The region of feasible values of $p_{ij}$ and $q_{ij}$ for a fixed $\epsilon_{ij}$ is a rhombus described in Proposition 2, which is easily verified. The optimum occurs at one of the corners, i.e., the corner when $p_{ij} = q_{ij}$. Thus, for each dyad $(i,j)$, we choose $p_{ij} = q_{ij} = 1 - \pi_{ij} = e^{\epsilon_{ij}}/(1 + e^{\epsilon_{ij}})$. This gives us $\epsilon_{ij} = \log \frac{(1-\pi_{ij})}{\pi_{ij}}$.

PROPOSITION 2. *Let $\epsilon_{ij}$ be fixed, then the region of feasible values for $p_{ij}$ and $q_{ij}$ is given by a rhombus defined by $LB(p_{ij}) \leq q_{ij} \leq UB(p_{ij})$ with*

$$LB(p_{ij}) = \begin{cases} 1 - e^{\epsilon_{ij}} p_{ij} & \text{if } 0 < p_{ij} < \frac{1}{1+e^{\epsilon_{ij}}} \\ e^{-\epsilon_{ij}}(1 - p_{ij}) & \text{if } \frac{1}{1+e^{\epsilon_{ij}}} < p_{ij} < 1 \end{cases},$$

$$UB(p_{ij}) = \begin{cases} 1 - e^{-\epsilon_{ij}} p_{ij} & \text{if } 0 < p_{ij} < \frac{e^{\epsilon_{ij}}}{1+e^{\epsilon}} \\ e^{\epsilon_{ij}}(1 - p_{ij}) & \text{if } \frac{e^{\epsilon_{ij}}}{1+e^{\epsilon_{ij}}} < p_{ij} < 1 \end{cases}.$$

The above result reveals an important limitation of measuring risk by the worst-case, as is done in DP, which is that the overall risk $\epsilon$ is always measured by the worst-case risk no matter if there maybe different risks for edges. Consider a situation where the risk of revealing the existence of an edge is more harmful than the non-existence of an edge. For instance, in a sexual partnership network, exposing a relationship between two individuals can be far more harmful than exposing that there is no relationship between them. However, DP does not recognize such a differential risk assigned to edges and non-edges: if the risk is measured by $\epsilon_{ij}$, then the optimal choice is to set $p_{ij} = q_{ij}$. Another situation where asymmetric risks may be useful, but DP focuses only on the overall risk, is when exposure of edges between certain types of nodes are considered more harmful than others; for example, in a sexual partnership network, edges between nodes of same sex may be more harmful than edges between nodes of different sex. This can be operationalized by setting different $\epsilon_{ij}$ levels for different pairs $(i,j)$, but, per Proposition 1, $\epsilon = \max_{ij} \epsilon_{ij}$, so to maintain a specific level of differential privacy every potential relationship must be treated as equally sensitive.

A justification often given for the requirement of measuring risk by worst-case in DP is that it allows for composition as described in Section 2: the risk cumulates over many different data releases in a controlled and predictable fashion (Dwork et al., 2006). The claim is that any non-worst case measure of risk may not compose in such a manner, but this is yet to be proven. Moreover, we are typically interested in releasing a small subset of synthetic networks for public use that would allow a wider range of statistical analysis than interactive data releases, thus limiting many more future data releases from the same dataset that could lead to quicker accumulation of overall risk.

### 4.2. Beyond Worst-Case Risk

The worst-case privacy risk of Algorithm 1, since it satisfies EDP, as measured by $\epsilon$ is determined by the most "revealing" dyad $(i,j)$, i.e., any dyad $(i^*, j^*)$ that achieves the maximum, has the highest $\epsilon_{ij}$ in Proposition 1. On the other hand, with our method, if we deem the disclosure of one set of dyads to be more harming than other, we can define a different risk measure for groups of dyads by specifying different values of $\epsilon$ for such groups. Consider partitioning the nodes into $K$ groups labeled by $k = 1, \ldots, K$. We can limit the privacy risk of dyads between nodes of groups $k_i$ and $k_j$ by specifying a $K \times K$ matrix of $\epsilon$ values. The $(k_i, k_j)$ entry of this matrix specifies the maximum tolerable privacy risk of dyads between nodes in group $k_i$ and $k_j$. The worst-case risk will still be determined by the maximum of all the $\epsilon_{k_i, k_j}$. The key point here is that having only a one number risk summary may not always be helpful, and one must be able to design mechanisms with different risks for different groups which is what we are able to do.

In practice, it may be acceptable to increase the risk of some dyads while decreasing the risk of others, in order to obtain more utility. It is important to note that the choice of risk should depend only on publicly available information. The choice of risk cannot depend on the existence of an edge in the network or the total number of edges between a group of nodes, but as in our framework, can depend on the attributes of the nodes as this information is assumed to be public. For example, one may deem that the re-identification of ties between nodes of same gender in a sexual network to be more devastating to the participants when compared to ties between different gender. In such a case, we may assign a lower value of $\epsilon$ (lower risk) for dyads between nodes of same sex, and a higher value of $\epsilon$ for all other dyads. Note that the overall worst-case risk is still determined by the largest $\epsilon$, but this setup allows one to take different risks into account. We use this strategy in the Enron e-mail case study in Section 6.

## 5. Likelihood based inference of ERGMs from randomized response

Exponential-family random graph models (ERGMs) (Wasserman and Pattison, 1996, and others) express the probability of a network $x \in \mathcal{X}$ as an exponential family:

$$P_{\boldsymbol{\theta}}(X = x) = \frac{\exp\{\boldsymbol{\theta} \cdot \boldsymbol{g}(x)\}}{c(\boldsymbol{\theta}, \mathcal{X})}, \quad x \in \mathcal{X}, \tag{1}$$

with $\boldsymbol{\theta} \in \Theta \subseteq \mathbb{R}^q$ a vector of parameters, $\boldsymbol{g}(x)$ a vector of sufficient statistics typically embodying the features of the social process that are of interest or are believed to be affecting the network structure, and $c(\boldsymbol{\theta}, \mathcal{X})$ is the normalizing constant given by

$$c(\boldsymbol{\theta}, \mathcal{X}) = \sum_{x \in \mathcal{X}} \exp\{\boldsymbol{\theta} \cdot \boldsymbol{g}(x)\}. \tag{2}$$

When $\boldsymbol{g}(x)$ can be decomposed into a summation over the network's dyads, i.e., $\boldsymbol{g}(x) = \sum_{i,j} x_{i,j} g_{i,j}$ for some covariate vector $g_{i,j}$, Model 1 becomes a logistic regression with the dyads as responses. Such a decomposition of the sufficient statistics (i.e., $\boldsymbol{g}(x)$) can be used to model a large variety of effects, including propinquity, homophily on observed attributes, and effects of actor attributes on gregariousness and attractiveness. However, substantively important effects like propensity towards monogamy in sexual partnership networks and triadic (friend-of-a-friend) effects in friendship networks cannot be modeled, and one needs to include sufficient statistics that induce *dyadic dependence*.

Under dyadic dependence, even when $x$ is fully observed (i.e., no privacy mechanism), it is a challenge to find the maximum likelihood estimate (MLE) of $\boldsymbol{\theta}$, because the normalizing constant $c(\boldsymbol{\theta}, \mathcal{X})$ given by (2) is an intractable sum over all ($2^{n(n-1)/2}$ for undirected) possible graphs in $\mathcal{X}$. Early efforts were limited to pseudolikelihood of Strauss and Ikeda (1990), but with availability of computing power, more accurate simulation-based methods were applied to the problem, first Robbins–Monro (Robbins and Monro, 1951) by Snijders (2002), then Monte-Carlo MLE (Geyer and Thompson, 1992) by Hunter and Handcock (2006). The latter algorithm starts with an initial guess $\boldsymbol{\theta}_0 \in \Theta$ and sets up a likelihood ratio between a candidate guess $\boldsymbol{\theta}$ near $\boldsymbol{\theta}_0$ and $\boldsymbol{\theta}_0$ itself,

$$\frac{L(\boldsymbol{\theta}, \mathcal{X})}{L(\boldsymbol{\theta}_0, \mathcal{X})} = \frac{\exp\{\boldsymbol{\theta} \cdot \boldsymbol{g}(x)\}/c(\boldsymbol{\theta}, \mathcal{X})}{\exp\{\boldsymbol{\theta}_0 \cdot \boldsymbol{g}(x)\}/c(\boldsymbol{\theta}_0, \mathcal{X})} = \exp\{(\boldsymbol{\theta} - \boldsymbol{\theta}_0) \cdot \boldsymbol{g}(x)\}\frac{c(\boldsymbol{\theta}_0, \mathcal{X})}{c(\boldsymbol{\theta}, \mathcal{X})},$$

then uses a sample under $\boldsymbol{\theta}_0$ to approximate the ratio $c(\boldsymbol{\theta}, \mathcal{X})/c(\boldsymbol{\theta}_0, \mathcal{X})$ by observing that

$$\begin{aligned}\frac{c(\boldsymbol{\theta}, \mathcal{X})}{c(\boldsymbol{\theta}_0, \mathcal{X})} &= \sum_{x' \in \mathcal{X}} \frac{\exp\{\boldsymbol{\theta} \cdot \boldsymbol{g}(x')\}}{c(\boldsymbol{\theta}_0, \mathcal{X})} = \sum_{x' \in \mathcal{X}} \exp\{(\boldsymbol{\theta} - \boldsymbol{\theta}_0) \cdot \boldsymbol{g}(x')\}\frac{\exp\{\boldsymbol{\theta}_0 \cdot \boldsymbol{g}(x')\}}{c(\boldsymbol{\theta}_0, \mathcal{X})} \\ &= E_{\boldsymbol{\theta}_0}[\exp\{(\boldsymbol{\theta} - \boldsymbol{\theta}_0) \cdot \boldsymbol{g}(x)\}] \approx \frac{1}{M}\sum_{i=1}^{M} \exp\{(\boldsymbol{\theta} - \boldsymbol{\theta}_0) \cdot \boldsymbol{g}(X_i)\}, \end{aligned} \quad (3)$$

for $X_1, X_2, \ldots, X_M$ a sample of $M$ realizations from the model at $\boldsymbol{\theta}_0$, simulated using MCMC (Snijders, 2002; Morris et al., 2008, for example). Maximizing the likelihood ratio with respect to $\boldsymbol{\theta}$ to obtain the next guess $\boldsymbol{\theta}_1$, simulating from $\boldsymbol{\theta}_1$, and repeating the process until convergence yields the MLE $\hat{\boldsymbol{\theta}}$.

Handcock et al. (2010) extended the above algorithm to the case where some dyads were unobserved—missing at random—and their approach can, in turn, be extended to private network data. Given a private network $y$ obtained by drawing one realization from $P_{\boldsymbol{\gamma}}(Y = y | X = x)$, simply maximizing $\boldsymbol{\theta}$ for $P_{\boldsymbol{\theta}}(X = y)$ can produce incorrect results (Karwa et al., 2014). Hence one must use the face-value likelihood $L_{y,\boldsymbol{\gamma}}(\boldsymbol{\theta})$, which sums over all possible true networks $x$ that could have produced $y$ via the privacy mechanism:

$$L_{y,\boldsymbol{\gamma}}(\boldsymbol{\theta}) = P_{\boldsymbol{\theta},\boldsymbol{\gamma}}(Y = y) = \sum_{x \in \mathcal{X}} P_{\boldsymbol{\theta},\boldsymbol{\gamma}}(Y = y \wedge X = x) = \sum_{x \in \mathcal{X}} P_{\boldsymbol{\theta}}(X = x) P_{\boldsymbol{\gamma}}(Y = y | X = x).$$

In case of the randomized response mechanism of Algorithm 1, $\boldsymbol{\gamma}$ is the collection of probabilities used for perturbing the dyads, i.e., $\boldsymbol{\gamma} = \{p_{ij}, q_{ij}\}$.

Now, consider the likelihood ratio of $\boldsymbol{\theta}$ with respect to some initial configuration $\boldsymbol{\theta}_0$:

$$\begin{aligned}\frac{L_{y,\boldsymbol{\gamma}}(\boldsymbol{\theta})}{L_{y,\boldsymbol{\gamma}}(\boldsymbol{\theta}_0)} &= \frac{\sum_{x \in \mathcal{X}} P_{\boldsymbol{\theta}}(X = x) P_{\boldsymbol{\gamma}}(Y = y | X = x)}{\sum_{x \in \mathcal{X}} P_{\boldsymbol{\theta}_0}(X = x) P_{\boldsymbol{\gamma}}(Y = y | X = x)} = \frac{\sum_{x \in \mathcal{X}} \frac{\exp\{\boldsymbol{\theta} \cdot \boldsymbol{g}(x)\}}{c(\boldsymbol{\theta}, \mathcal{X})} P_{\boldsymbol{\gamma}}(Y = y | X = x)}{\sum_{x \in \mathcal{X}} \frac{\exp\{\boldsymbol{\theta}_0 \cdot \boldsymbol{g}(x)\}}{c(\boldsymbol{\theta}_0, \mathcal{X})} P_{\boldsymbol{\gamma}}(Y = y | X = x)} \\ &= \frac{c(\boldsymbol{\theta}_0, \mathcal{X})}{c(\boldsymbol{\theta}, \mathcal{X})} \sum_{x \in \mathcal{X}} \exp\{(\boldsymbol{\theta} - \boldsymbol{\theta}_0) \cdot \boldsymbol{g}(x)\} \frac{\exp\{\boldsymbol{\theta}_0 \cdot \boldsymbol{g}(x)\} P_{\boldsymbol{\gamma}}(Y = y | X = x)}{\sum_{x' \in \mathcal{X}} \exp\{\boldsymbol{\theta}_0 \cdot \boldsymbol{g}(x')\} P_{\boldsymbol{\gamma}}(Y = y | X = x')} \\ &= \frac{c(\boldsymbol{\theta}_0, \mathcal{X})}{c(\boldsymbol{\theta}, \mathcal{X})} E_{\boldsymbol{\theta}_0, \boldsymbol{\gamma}}[\exp\{(\boldsymbol{\theta} - \boldsymbol{\theta}_0) \cdot \boldsymbol{g}(x)\} | Y = y]. \end{aligned}$$

In the above expression, the ratio of normalizing constants can be approximated as in (3), while the conditional expectation can, as before, be approximated using MCMC. Karwa et al. (2014) did so by re-weighting the sample drawn to estimate $c(\boldsymbol{\theta}, \mathcal{X})/c(\boldsymbol{\theta}_0, \mathcal{X})$ by $P_{\boldsymbol{\gamma}}(Y = y|X = X_i)$, which is similar to a weighted EM based approach of Woo and Slavković (2012). The approach of Karwa et al. (2014) is very costly, however, because the weights $P_{\boldsymbol{\gamma}}(Y = y|X = X_i)$ are likely to be concentrated among the small number of $X_i$ closest to $y$, requiring a huge $M$ for an adequate precision. We propose to instead draw a second MCMC sample, from $P_{\boldsymbol{\theta}_0,\boldsymbol{\gamma}}(X = x|Y = y)$, using a Metropolis algorithm that proposes symmetrically over $X$ and accepts with probability

$$\min\left\{1, \frac{\exp\{\boldsymbol{\theta}_0 \cdot \boldsymbol{g}(x^\star)\} P_{\boldsymbol{\gamma}}(Y = y|X = x^\star)}{\exp\{\boldsymbol{\theta}_0 \cdot \boldsymbol{g}(x)\} P_{\boldsymbol{\gamma}}(Y = y|X = x)}\right\} = \min\left(1, \exp[\boldsymbol{\theta}_0 \cdot \{\boldsymbol{g}(x^\star) - \boldsymbol{g}(x)\}] \frac{P_{\boldsymbol{\gamma}}(Y = y|X = x^\star)}{P_{\boldsymbol{\gamma}}(Y = y|X = x)}\right),$$

or a similar Metropolis–Hastings algorithm with an asymmetric proposal. Given such a sample, $X'_1, X'_2, \ldots, X'_M$,

$$\frac{L_{y,\boldsymbol{\gamma}}(\boldsymbol{\theta})}{L_{y,\boldsymbol{\gamma}}(\boldsymbol{\theta}_0)} \approx \frac{\frac{1}{M}\sum_{i=1}^M \exp\{(\boldsymbol{\theta} - \boldsymbol{\theta}_0) \cdot \boldsymbol{g}(X'_i)\}}{\frac{1}{M}\sum_{i=1}^M \exp\{(\boldsymbol{\theta} - \boldsymbol{\theta}_0) \cdot \boldsymbol{g}(X_i)\}}. \tag{4}$$

The computing cost of fitting a given ERGM to a private sample $y$ is therefore on the order of twice the cost of fitting the same ERGM to the fully observed network $x$. As described, the two-sample approach does not remove the requirement that it be possible to compute probabilities $P_{\boldsymbol{\gamma}}(Y = y|X = x)$ (or at least their ratio $P_{\boldsymbol{\gamma}}(Y = y|X = x^\star)/P_{\boldsymbol{\gamma}}(Y = y|X = x)$) in a closed form: the parameters of the privacy mechanism still need to be known. The standard errors can be obtained by inverting the negative Hessian of the face-value log-likelihood, and twice differentiating the logarithm of (4) with respect to $\boldsymbol{\theta}$ suggests estimating the latter by applying the formula of Hunter and Handcock (2006, eq. 3.5) to the constrained and the unconstrained samples and taking the difference, that is,

$$\hat{I}_{y,\boldsymbol{\gamma}}(\hat{\boldsymbol{\theta}}) \approx \left\{\overline{\boldsymbol{g}(X_i)\boldsymbol{g}(X_i)^\top} - \overline{\boldsymbol{g}(X_i)}\,\overline{\boldsymbol{g}(X_i)}^\top\right\} - \left\{\overline{\boldsymbol{g}(X'_i)\boldsymbol{g}(X'_i)^\top} - \overline{\boldsymbol{g}(X'_i)}\,\overline{\boldsymbol{g}(X'_i)}^\top\right\},$$

where $\overline{f(X_i)} = M^{-1}\sum_{i=1}^M w_i f(X_i)$ some function $f(\cdot)$,

$$w_i = \exp\left\{(\hat{\boldsymbol{\theta}} - \boldsymbol{\theta}_{\text{last}}) \cdot \boldsymbol{g}(X_i)\right\} / \sum_{i=1}^M \exp\left\{(\hat{\boldsymbol{\theta}} - \boldsymbol{\theta}_{\text{last}}) \cdot \boldsymbol{g}(X_i)\right\},$$

and analogously for $w'_i$ and $\overline{f(X'_i)}$. We implemented this inference technique as an enhancement to the `ergm` package (Hunter et al., 2008; Handcock et al., 2015), which we intend to make available for public use.

## 6. Case study: Enron e-mail data

In this section we study the private release and analysis of the *Enron* dataset discussed in Section 1 by applying the randomized response mechanism with the likelihood based inference developed in sections 3 and 5. A similar study of the *Lazega* and the *Teenage friendship* data can be found in the supplement, (Karwa et al., 2016).

The goal of these case studies is to demonstrate how one can share differentially private synthetic networks for valid statistical analyses. We synthesize and "release" networks by generating copies of the original network using Algorithm 1. By assuming that the potential user would only have access to the released synthetic network and the knowledge of the privacy mechanism, we evaluate how well can a researcher replicate an analysis performed on the original data and obtain inferences similar to those using the original network. The synthetic network can be analyzed via two methods: the *Naive* where one ignores the privacy mechanism and analyzes the synthetic network as if it was the original network, and the *Missing Data* from Section 5 where one models the privacy-preserving mechanism explicitly and parameter estimates are obtained by maximizing the missing data likelihood. Recall, we assume that the attribute information associated with the nodes is publicly available, for reasons explained in Section 1.

### 6.1. Evaluating accuracy

We evaluate the accuracy of the estimates by using Kullback–Leibler (KL) divergence, mean squared error (MSE) and the bias. KL divergence measures the distance between two distributions on networks – the first one implied by estimates $\hat{\boldsymbol{\theta}}_y$ obtained from the synthetic network $y$ and the second one implied by the estimates $\hat{\boldsymbol{\theta}}_x$ from the original network $x$. A smaller value of KL suggests that the estimated models are close. To obtain insight into the accuracy of individual parameter estimates, we measure their bias and MSE. The bias is defined as $E[\hat{\boldsymbol{\theta}}_y] - \hat{\boldsymbol{\theta}}_x$ and the MSE is given by $E[(\hat{\boldsymbol{\theta}}_x - \hat{\boldsymbol{\theta}}_y)^2]$.

The KL divergence between the two distributions:

$$KL(\hat{\boldsymbol{\theta}}_x, \hat{\boldsymbol{\theta}}_y) = E_{\hat{\boldsymbol{\theta}}_x}\left[\log \frac{P_{\hat{\boldsymbol{\theta}}_x}(X)}{P_{\hat{\boldsymbol{\theta}}_y}(X)}\right] = \sum_{x \in \mathcal{X}} \log\left(\frac{P_{\hat{\boldsymbol{\theta}}_x}(x)}{P_{\hat{\boldsymbol{\theta}}_y}(x)}\right) P_{\hat{\boldsymbol{\theta}}_x}(x)$$

$$= (\hat{\boldsymbol{\theta}}_x - \hat{\boldsymbol{\theta}}_y) \cdot \boldsymbol{g}(x) + \log \frac{c(\hat{\boldsymbol{\theta}}_y)}{c(\hat{\boldsymbol{\theta}}_x)} = \log \frac{L_x(\hat{\boldsymbol{\theta}}_x)}{L_x(\hat{\boldsymbol{\theta}}_y)}.$$

This log-likelihood-ratio can be estimated using the MCMC techniques described in Section 5 and by Handcock et al. (2010).

We consider two ways of measuring risks for releasing synthetic networks. For both, for each value of privacy risk, we release $B = 20$ synthetic networks using Algorithm 1 and estimate their ERGM parameters. In *Method 1*, we assign different privacy risks for different types of edges, when it is believed that revealing certain types of ties, given the nodal information, will have higher privacy risks associated with them. The choice of which ties are riskier if revealed of course depends on the application. In *Method 2*, we assign equal privacy risks for all dyads by setting $p_{ij} = q_{ij} = 1 - \pi$, where $1 - \pi$ is the probability of retaining an edge (or non-edge) and $\pi$ the probability of perturbing an edge (or non-edge). For understanding the risk-utility trade-off, we use a range of values of $\pi$ in Algorithm 1, and plot the KL divergence between the estimates obtained using the synthetic networks and the estimates from the original network. The horizontal axis denotes the privacy risk as measured by the probability of perturbing an edge $\pi$ in %. Note that larger values of $\pi$ imply weaker privacy. The vertical axis denotes utility of the synthetic networks as measured by the KL divergence on *log* scale. Larger values of KL divergence correspond to lower utility; e.g., see Figure 1b.

### 6.2. Data and Model

We consider a large subset of Enron email network (Cohen, 2009), and use a version of data with email communications aggregated over the entire period of study between 1998 and 2002 (Perry and Wolfe, 2013; Zhou et al., 2007). Viewing a mass-mailing as representing a weaker social link than a personal message, we weight each message by the inverse of the number of its recipients and consider there to be a link from $i$ to $j$ if the sum of message weights with $i$ as the sender and with $j$ among its recipients is greater than or equal to 2. Note that thresholding a valued dataset can produce misleading results (Butts, 2009), and privacy for valued network models is subject of ongoing work.

In our analysis, we consider the effects of exogenous attributes of employees (gender, seniority, and department) and endogenous effects (mutuality). For the former, we ask if those equally senior are more likely to communicate than those with different levels of seniority, as well as whether senior employees are more likely to e-mail junior employees, rather than the other way around, and similarly for the two genders; and we measure how much the departments' overall propensities to communicate vary, and how much more communication occurs within a department as opposed to between them.

The network consists of 156 nodes and 1310 egdes. There are three covariates associated with each node: *department* (Legal, Trading or Other), *gender* (Male or Female), *Seniority* (Senior, Junior). We consider an ERGM with endogenous effects modeled by the number of edges and mutuality, exogenoues effects modeled by the direct effects and the homophily effects of *Gender*, *Department* and *Seniority*. The direct effects are given by $\sum_{i,j} x_{ij} Z_i$, where $Z_i$ is the attribute of node $i$. The dyadic homophily effects are given by $\sum_{ij} x_{ij} I(Z_i = Z_j)$ where $Z_i$ is the attribute of node $i$. These terms capture the matches between the two nodes on the given attribute and measure the strength of ties between nodes of the same attribute type.

### 6.3. Results

We consider two methods for selecting perturbation probabilities: one that perturbs dyads uniformly and another that varies perturbation probabilities based on known actor attributes.

*Method 1:* We assign disclosure risks based on covariates, such that email exchanges between nodes from the Legal Department are deemed to be riskier than other emails, and hence are given more privacy protection, i.e. they have a higher probability of perturbation. We use two different values for the probability of perturbing a dyad: $\pi_1 = 4.8\%$ (i.e., privacy risk $\epsilon = 3$) if both nodes $i$ and $j$ belong to the Legal Department, and $\pi_2 = 0.25\%$, (i.e., privacy risk $\epsilon = 6$), otherwise. The overall privacy risk for any dyad is measured by larger $\epsilon$, 6 (higher privacy risk) in this case. However, the privacy risk for emails exchanged between nodes that belong to the Legal Department is 3 (lower privacy risk) and the privacy protection is higher. We compare this network release strategy with two others where all dyads are released either with $\pi = \pi_1$ or with $\pi = \pi_2$. Figure 1a shows the utility (as measured by KL divergence) for these three strategies; the utility is lowest (KL divergence is highest) when we use a perturbation probability of 4.8% which corresponds to the lowest privacy risk. However, if we allow for larger privacy risks for emails from the Legal Department ($\pi_1 = 4.8\%$ and $\pi_2 = 0.25\%$), we obtain an improved utility. Finally, if we increase the privacy risk for all emails ($1 - \pi = 0.0025$), we obtain marginally higher utility, but at the expense of reduced privacy. This analysis shows that assigning asymmetric privacy risks offers a better risk-utility trade-off.

*Method 2:* We assume the same the privacy risk for every dyad and set the probability of perturbing a dyad $1 - p_{ij} = 1 - q_{ij} = \pi$. Figures 1b, 1c and 1d show the results, and Figure 2 shows the original sub-network of randomly chosen 50 nodes with 3 synthetic sub-networks generated using $\pi = 2\%$. Notice the addition of fake ties and removal of existing ties. The synthetic sub-networks are visibly more dense: while 2% may seem like a tiny perturbation probability, consider that the original network has a density of 5.4%, which means that, effectively, more than a quarter of the ties in the network are now "fake". Figure 1b shows that the KL divergence between the private estimate and the non-private estimate increases as the probability of perturbing an edge increases, thus stronger privacy leads to reduced utility. However, note that since the vertical axis is in the log scale, as we increase privacy (move from left to right on the horizontal axis) the KL divergence of the naive estimates increases at a much faster rate when compared to the missing data estimates, especially for larger values of the perturbation probability (e.g., 1 to 3 percent). Thus for strong privacy protection, the missing data likelihood provides estimates that are closer to the non-private estimates. Uncertainty in model estimation is reflected by the variation in the KL divergence. We can see that for perturbation probability of 2% and 3%, the KL divergence of the missing data method shows much higher variation than the naive method, because for larger values of perturbation probability, there are a lot of fake edges that are added and the uncertainty in model estimation increases. On the other hand, the naive method incorrectly places

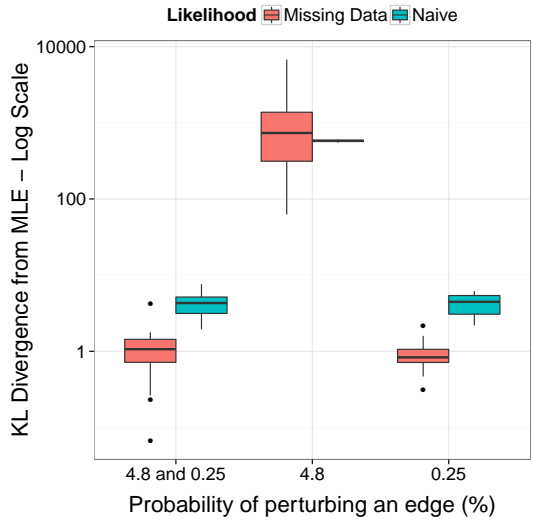
(a) KL Divergence - Method 1

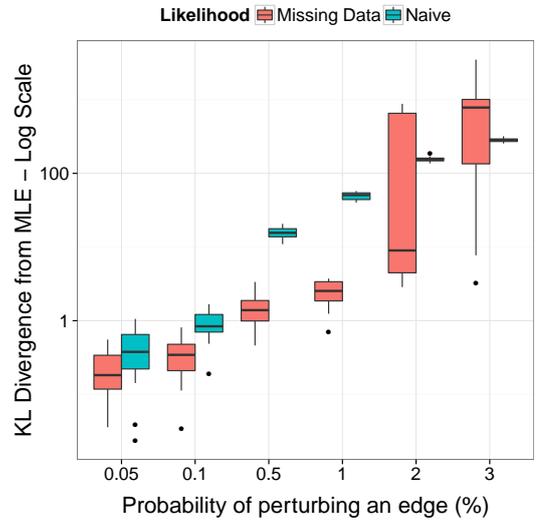
(b) KL Divergence - Method 2

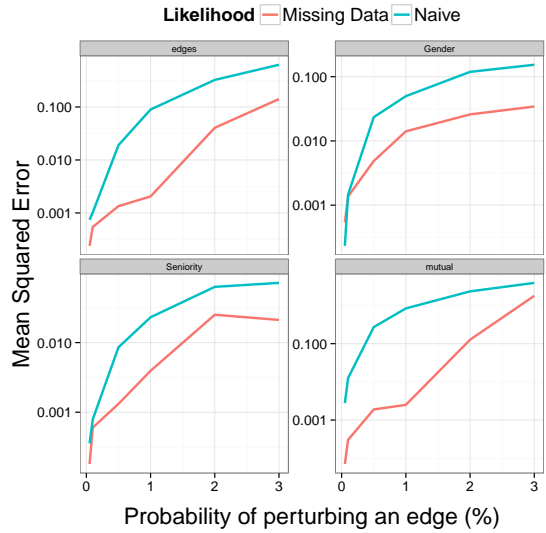
(c) Mean Squared Error

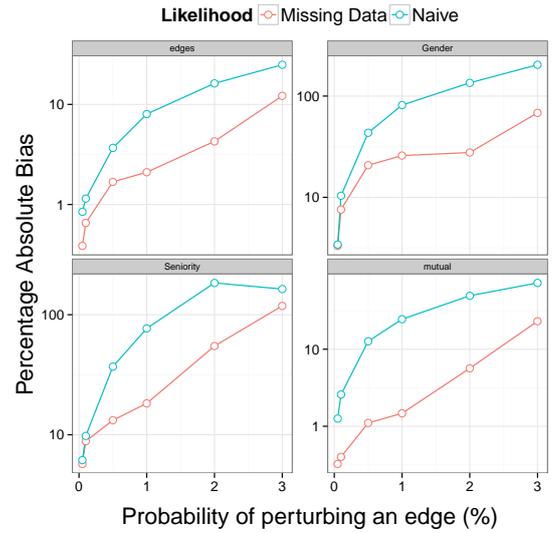
(d) Percentage Absolute Bias of parameters

**Fig. 1.** Enron data: (a) and (b) Box plots of KL divergence of models obtained using the synthetic networks. The horizontal axis represents the probability of perturbing an edge in % (higher corresponds to lower privacy risk). The vertical axis shows the KL divergence in log scale (higher value corresponds to lower utility). The red color (lower value of KL) represents the KL divergence between the missing data estimates the MLE obtained from the original data. The cyan line (higher value of KL) represents the KL divergence between the naive estimates and the MLE; (c) MSE; (d) Absolute Bias of parameter estimates.

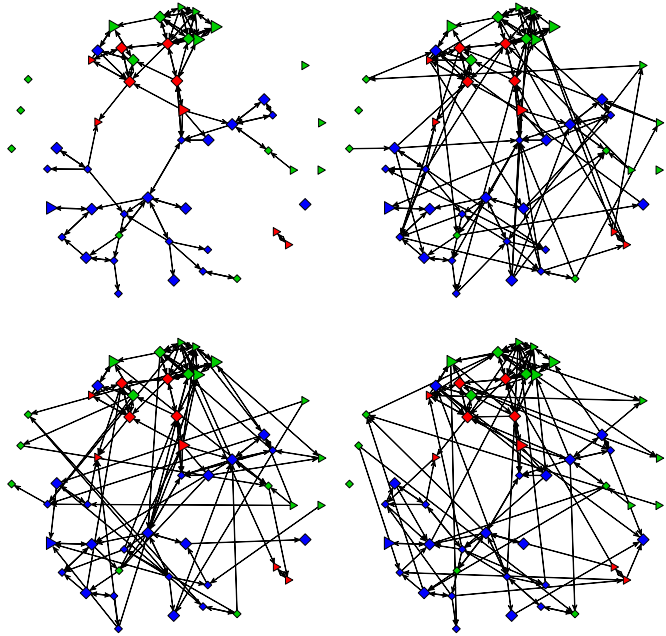

**Fig. 2.** 3 Synthetic copies of a random sub-network of 50 nodes of the Enron Network for 2% perturbation and the original sub-network (the top left corner) are plotted in the same coordinate system for ease of comparison.

a lot of confidence in the estimated model since it ignores the privacy mechanism.

For a more detailed evaluation, figures 1c and 1d show the MSE and the percentage absolute bias (with respect to the MLE from the original data) of both the missing data and naive parameter estimates obtained from the synthetic networks with varying levels of perturbation; overall, the missing data results are better, but not uniformly. Table 1 shows the mean parameter estimates for a 2% perturbation with their bias and the MSE. Here, the bias of the missing data estimates is smaller than of the naive estimates, but the MSE is typically larger. This is expected as the missing data estimates take into account the additional uncertain in the privacy mechanism.

Table 2 gives a typical (median, according to KL divergence) fits from the simulation, their standard errors, and parameter significance levels also for a 2% perturbation. Here, the naive estimator's standard errors are uniformly smaller compared to the those from the original data. This is likely because the perturbed network is more dense, resulting in higher variability of sufficient statistics under the model and thus higher

**Table 1.** Parameter estimates based on the original data (MLE) and the synthetic networks (Missing and Naive) obtained for the Enron Data for a 2% perturbation.

| Parameter | MLE | Missing Data | | | Naive Data | | |
| --- | --- | --- | --- | --- | --- | --- | --- |
| | | estimate | mse | bias | estimate | mse | bias |
| Edges | -3.98 | -3.70 | 0.21 | 0.29 | -3.40 | 0.34 | 0.58 |
| Node mix (Gender) | -0.61 | -0.42 | 0.09 | 0.20 | -0.27 | 0.13 | 0.34 |
| Node mix (Seniority) | -0.43 | -0.18 | 0.10 | 0.26 | -0.15 | 0.09 | 0.29 |
| Mutuality | 4.65 | 3.34 | 45.73 | -1.31 | 3.12 | 2.37 | -1.54 |
| Nodefactor Department (Other) | 0.05 | 0.05 | 0.01 | -0.01 | 0.04 | 0.00 | -0.01 |
| Nodefactor Department (Trading) | -0.15 | -0.12 | 0.01 | 0.03 | -0.11 | 0.00 | 0.04 |
| Homophiliy Department (Legal) | 1.50 | 1.16 | 0.54 | -0.34 | 1.42 | 0.01 | -0.08 |
| Homophiliy Department (Other) | 0.34 | 0.23 | 0.02 | -0.11 | 0.29 | 0.00 | -0.05 |
| Homophiliy Department (Trading) | 0.63 | 0.49 | 0.10 | -0.14 | 0.49 | 0.02 | -0.14 |
| Homophiliy Gender (Female) | 0.02 | 0.04 | 0.04 | 0.02 | 0.14 | 0.02 | 0.12 |
| Homophiliy Gender (Male) | -0.26 | -0.15 | 0.03 | 0.11 | -0.11 | 0.03 | 0.16 |
| Homophiliy Seniority (Junior) | -0.10 | -0.02 | 0.01 | 0.08 | 0.02 | 0.01 | 0.12 |
| Homophiliy Seniority (Senior) | 0.50 | 0.44 | 0.07 | -0.06 | 0.53 | 0.00 | 0.03 |

**Table 2.** Parameter estimates, std. errors, and significance of the original Enron data (MLE) and the models fit to synthetic networks with 2% perturbation using Missing Data MLE and Naive fit, all based on the median, according to KL-divergence.

| Parameter | MLE | Missing Data | Naive |
| --- | --- | --- | --- |
| Edges | $-3.982\ (0.160)^{***}$ | $-4.354\ (0.199)^{***}$ | $-3.471\ (0.130)^{***}$ |
| Mutuality | $4.654\ (0.116)^{***}$ | $4.510\ (0.165)^{***}$ | $3.070\ (0.088)^{***}$ |
| Homophily Seniority (Junior) | $-0.098\ (0.099)$ | $0.102\ (0.122)$ | $-0.024\ (0.075)$ |
| Homophily Seniority (Senior) | $0.500\ (0.097)^{***}$ | $0.701\ (0.119)^{***}$ | $0.492\ (0.071)^{***}$ |
| Node mix (Seniority) | $-0.434\ (0.162)^{**}$ | $-0.080\ (0.213)$ | $-0.211\ (0.105)^{*}$ |
| Homophily Gender (Female) | $0.022\ (0.114)$ | $0.211\ (0.141)$ | $0.211\ (0.094)^{*}$ |
| Homophily Gender (Male) | $-0.263\ (0.095)^{**}$ | $-0.085\ (0.126)$ | $-0.044\ (0.073)$ |
| Node mix (Gender) | $-0.615\ (0.170)^{***}$ | $-0.207\ (0.225)$ | $-0.174\ (0.114)$ |
| Nodefactor Department (Other) | $0.054\ (0.097)$ | $0.151\ (0.108)$ | $0.040\ (0.081)$ |
| Nodefactor Department (Trading) | $-0.147\ (0.084)$ | $-0.181\ (0.092)^{*}$ | $-0.021\ (0.076)$ |
| Homophily Department (Legal) | $1.499\ (0.134)^{***}$ | $1.556\ (0.148)^{***}$ | $1.487\ (0.126)^{***}$ |
| Homophily Department (Other) | $0.336\ (0.127)^{**}$ | $0.224\ (0.134)$ | $0.352\ (0.111)^{**}$ |
| Homophily Department (Trading) | $0.633\ (0.128)^{***}$ | $0.737\ (0.145)^{***}$ | $0.397\ (0.113)^{***}$ |

Significance levels: $0.05 \geq^{*} > 0.01 \geq^{**} > 0.001 \geq^{***}$

estimate of information (Hunter and Handcock, 2006, eq. 3.5) and smaller standard errors. However, the missing data MLE's standard errors are uniformly larger because they reflect (correctly) the additional uncertainty introduced by the perturbation. This shows that the overall inference even from the highly perturbed networks is still valid given our methods, albeit with some loss of statistical power. While the missing data MLE reliably detects mutuality, homophily between senior employees, and other strong ($P$-value $\leq 0.001$) effects, other effects, homophily in the "Other" department or the tendency of males to e-mail females more often than otherwise, are lost. For less perturbation, some power would be regained, and thus the data curator will need to make a choice on a desirable risk-utility trade-off.

## 7. Conclusions

Motivated by a growing availability of network data combined with growing concerns about privacy, we describe a framework for sharing relational data that not only preserves the privacy of individual relationships in a quantifiable manner, but also allows for meaningful inferences in estimating the popular exponential-family random graph models. The randomized response scheme we propose is simple yet effective, and quantifiable via the Edge Differential Privacy framework that measures privacy risk in terms of a worst-case disclosure. We performed a case study to evaluate how well the proposed approach works at a variety of privacy levels, and we include in the supplement additional case studies, with networks differing in size, structure, topology and associated covariate information, and demonstrate its usefulness in addressing the realistic challenge of simultaneously maintaining the privacy of sensitive relations in the network and sharing of the network data that would support valid statistical inference. Our analyses show that the proposed approach leads to estimates much closer to those obtained for a full network than those obtained by ignoring the privacy mechanism. We can replicate the original analyses using synthetic networks, but we need to model the privacy mechanism that generated the network explicitly and use the missing data likelihood. Although we advocate the use of missing data and MCMC techniques by analysts who use data obtained from a differentially private mechanism, or more general privacy-preserving mechanisms, they can also be used by data curators to release synthetic graphs for performing preliminary analysis of other models.

We are not advocating the use of such synthetic datasets for performing publication quality analysis, although in some cases, the synthetic data may suffice. Such synthetic datasets should be seen as a way to provide easy access to datasets that otherwise would never be made available or would require years before access is granted. The synthetic datasets can be used to perform qualitative, exploratory or preliminary data analyses. If the dataset seems useful or a more detailed analysis is required, one must obtain access to the original data through contractual obligations and other means.

A key advantage of this method in relation to other proposed methods for private network release is that the released private synthetic graphs preserve the actual relations and not just sufficient statistics, so our technique allows us to find MLE of any ERGM that could have been fitted to the original network, at a modest computational cost. The only alternative computational approach known to us, the Bayesian exchange algorithm, requires an MCMC sample of network realizations *for each* MCMC draw of $\theta$ from the posterior (MCMC within MCMC) which vastly increases the computational cost, when our approach merely doubles it, with the two samples able to be run in parallel.

In addition, having estimated $\hat{\theta}_y$ from the perturbed graph, we can simulate from the conditional distribution $X|\hat{\theta}_y, y$ ($\propto P_\gamma(Y = y|X = x)P_{\hat{\theta}}(X = x)$) of possible graphs $x$ from which $y$ could have plausibly come. For example, if $x$ exhibited strong homophily on some actor attribute, $y$, which has had false ties added at random, would exhibit weaker homophily. The amount of homophily in $x$ could be estimated by the ERGM using our technique, and graphs simulated from $X|\hat{\theta}_y, y$ would retain most of the relations in $y$, but "clean" many of the false ties inconsistent with the model, as demonstrated by the case studies.

We have used differential privacy as our measure of protection, but this approach, while it provides strong guarantees has substantive limitations, as discussed in Section 4. For example, we distinguished $1 - p_{ij}$, the probability of hiding a tie, from $1 - q_{ij}$, the probability of creating a false tie. Our inferential framework handles this seamlessly, and this distinction is important if the ties reflect socially or legally stigmatized relationships.

In that case, one might want to set $1 - q_{ij}$ to a relatively high value in order to create deniability for actors with such relationships. However, setting $1 - p_{ij} > 0$ would reduce utility to little gain of privacy. Thus, one might set $p_{ij} = 1$, but then Proposition 1 gives $\epsilon = \infty$. This suggests that this measure is too crude to assess disclosure risk when there is an assymmetry in the consequences of a tie as opposed to a non-tie being exposed.

Lastly, we assume that while the relationships are sensitive, the exogenous individual attributes such as gender are not: they can be released completely and without noise. This is a limitation inherent in ERGMs, which treat them as fixed and known covariates. The exponential-family random network models introduced by (Fellows and Handcock, 2012) propose to model relations and actor attributes jointly in an exponential family framework. If actor attributes are perturbed as well with a known probability, our inferential approach should be directly applicable, and this is subject for ongoing research.

## 8. Supplementary Material

### 8.1. The Lazega Collaboration Network

The Lazega dataset records the collaborative working relations between partners in a New England firm (Lazega, 2001) and have been analyzed in Snijders et al. (2006) and Hunter and Handcock (2006). Following Handcock et al. (2010), we focus on the undirected network where an edge between two partners exist if they collaborate with each other. The network consists of $n = 36$ nodes and 115 edges (see Figure 3a), and is supplemented by four attributes: *Seniority* (the rank number of chronological entry into the firm), *Practice* (litigation $=0$ and corporate law $= 1$), *Gender* (3 out of the 36 lawyers are female), and *Office* (three different offices based in three different cities). While there are no obvious privacy concerns with these data as with the two other case studies, one can imagine a scenario where certain partnerships would be exclusive given the nodal information such as gender. Our goal is to assess the effect of fitting the same model with privacy constraints and evaluate if we can replicate the findings, in particular the parameter estimates, using only synthetic networks released by Algorithm 1 in the main paper.

In comparison to Enron case study, here we explore an undirected network and fit an ERGM with 7 parameters: two parameters corresponding to the direct effects of *Seniority* and *Practice* and three parameters corresponding to the homophily effects of *Practice*, *Gender* and *Office*, and two parameters *Edges* (the number of edges in the network) and $GWESP$ to capture the network structure. The $GWESP$ term represents geometrically weighted edgewise shared partner distributions, a measure of transitivity structure in the network (Snijders et al., 2006). This particular form of ERGM was used by Handcock et al. (2010), and an equivalent form was used in Hunter and Handcock (2006) who found that the model provides an adequate fit to the data and can be successfully used to describe the structure of the Lazega dataset, that is explain the observed patterns of collaborative ties as a function of nodal and relational attributes.

Here, we create synthetic networks only using *Method 2*, i.e., the same probability of retaining an edge (or non-edge) for all dyads. Figure 3a shows plots of the original network and 8 randomly chosen synthetic networks for a perturbation probability $\pi = 2\%$.

For understanding the risk-utility trade-off, we use a range of $\pi$ values, i.e., the probability of perturbing an edge (in percentage), to be $0.1, 0.2, 0.4, 0.5, 1.0, 2.0, 5.0, 10.0$. Figure 3b shows that ignoring the privacy mechanism when analyzing the synthetic network leads to a much lower utility while using the missing data method leads to improved utility. Figures 3c and 3d with the MSE and the percentage absolute bias of the parameter estimates for the Lazega data, in comparison to the Enron data ones, show more stability and convincing evidence that as $\pi$ (probability of perturbing an edge) increases, the MSE and the bias are increase and second, as expected, the estimates based on the missing data method for all parameters have smaller bias and MSE than the naive estimates. The bias for the *missing data* estimates is always lower than for the *naive* estimates. However, when $\pi$ is close to 0, the MSE of the missing data estimates spikes and is larger than that of the naive estimates. This shows that the variance of the missing data estimates is much larger for values of $\pi$ close to 0. When $\pi$ is close to 0, the synthetic network is not very different from the original network, so the noise due to the MCMC estimation in missing data method increases the variance.

Table 8.1 shows the mean parameter estimates, bias and the MSE of the estimates

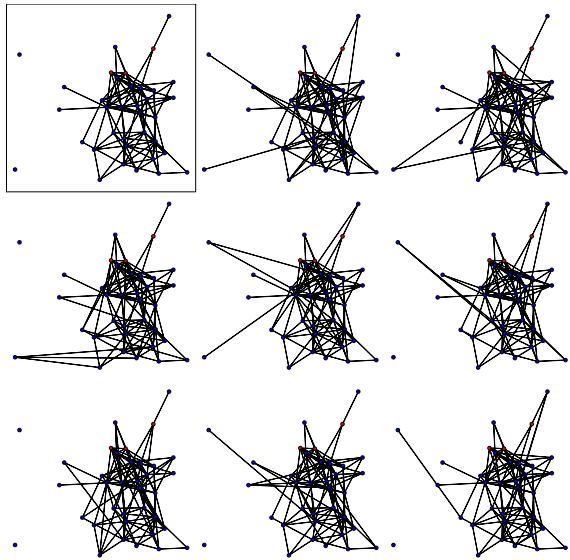

(a) 8 Synthetic copies of the Lazega Network using $\pi = 2\%$

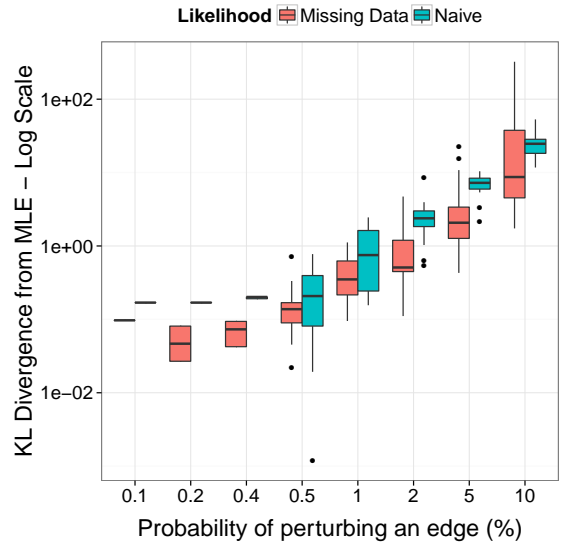

(b) KL divergence

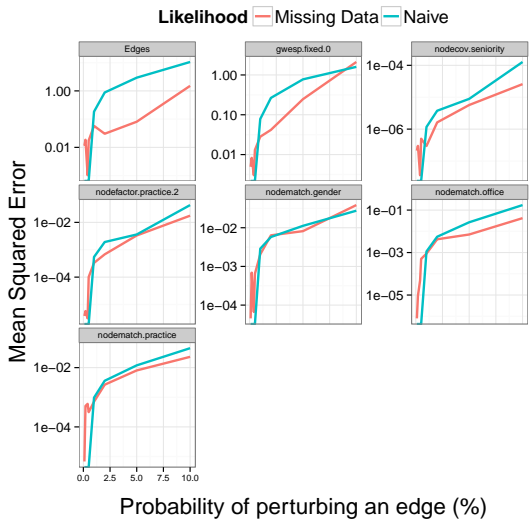

(c) Mean Squared Error

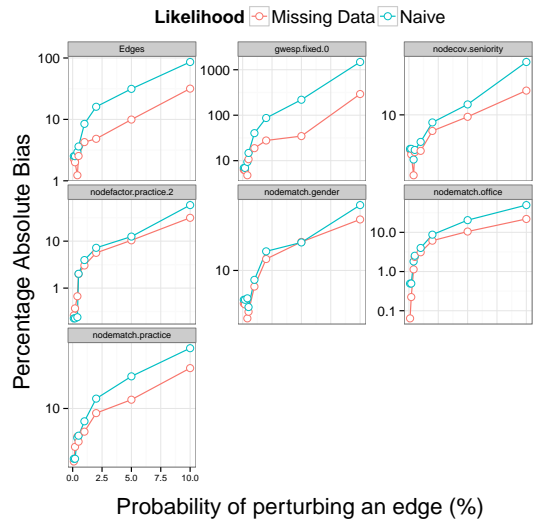

(d) Percentage Absolute Bias of parameters

**Fig. 3.** (a) Original and Synthetic copies of the Lazega Network (b) Box plots of log KL divergence of models obtained using the synthetic networks. The $x$-axis represents the probability of perturbing an edge $\pi$ (higher corresponds to lower privacy risk). The $y$-axis shows the KL divergence in log scale (higher corresponds to lower utility). The red color (lower value of KL) represents the KL divergence between the missing data estimates the MLE obtained from the original data. The cyan line (higher value of KL) represents the log KL divergence between the naive estimates and the MLE. (b) and (c) MSE and Percentage Absolute Bias of the parameter estimates obtained using the synthetic networks of the Lazega Data.

**Table 3.** Table showing the parameter estimates based on the original data (MLE) and the synthetic networks (Missing and Naive) obtained for the Lazega Data for $\pi = 2\%$.

| Parameter | MLE | Missing Data | | | Naive | | |
|---|---|---|---|---|---|---|---|
| | | Estimate | MSE | Bias | Estimate | MSE | Bias |
| Edges | -7.33 | -7.32 | 0.21 | 0.01 | -6.33 | 1.1 | 1 |
| gwesp.fixed.0 | 1.48 | 1.52 | 0.2 | 0.03 | 0.89 | 0.42 | -0.6 |
| nodecov.seniority | 0.04 | 0.04 | 0 | 0 | 0.03 | 0 | 0 |
| nodefactor.practice.2 | 0.75 | 0.74 | 0 | 0 | 0.72 | 0 | -0.03 |
| nodematch.gender | 0.93 | 0.89 | 0.02 | -0.04 | 0.86 | 0.02 | -0.07 |
| nodematch.office | 1.41 | 1.4 | 0.01 | 0 | 1.32 | 0.02 | -0.09 |
| nodematch.practice | 0.84 | 0.81 | 0.01 | -0.03 | 0.75 | 0.01 | -0.09 |

obtained by using synthetic networks generated by setting $\pi = 2\%$, which corresponds to $\epsilon = 3.89$, and further supports the claim that the missing data estimates have a very small empirical bias, i.e., they are close to the MLEs based on the original network. Note that the estimates of the structural parameters and the homophily effects are more biased when compared to the main effects of the nodes, but the overall inference, most importantly, would be the same if one is to use the original data.

The results of this case study show that we can replicate the analysis using synthetic networks, but we need to model the mechanism that generated the network explicitly and use the missing data likelihood. For extremely small values of perturbing an edge, it appears that the missing data estimates only increase the variance, as the estimates are already unbiased and close to the MLE.

### 8.2. Teenage friendship and substance use data

The teenage friendship network, which is a part of the data collected in the Teenage Friends and Lifestyle Study (Michell and Amos, 1997; Pearson and Michell, 2000), records friendships and substance use for a cohort of students in a school in Scotland. We use an excerpt of 50 adolescent girls made available online in the *Siena* package (Siena, 2014) where the network consists of 50 nodes and 39 edges. There are four covariates associated with each node: *Drug usage* (yes or no), *Smoking status* (yes or no),

*Alcohol usage*, (regular or irregular) and *Sport activity* (regular or irregular). As before, we assume that the attributes associated with each node are available publicly, and we aim to protect the relationship information in the network and perform inference by fitting an ERGM to differentially private synthetic networks. Figure 4 shows the original friendship network, and a set of synthetic networks, to which we fit the following ERGM:

$$P_\theta(X) \propto \exp\{\theta_1 \text{edges} + \theta_2 \text{gwesp} + \theta_3 \text{popularity} + \theta_4 \text{drug} + \theta_5 \text{sport} + \theta_6 \text{smoke} + \theta_7 \text{sport:alchohol}\}. \tag{5}$$

The first three terms capture the network structure of the graph (the number of edges, transitive effects and the degree distribution), the next three terms represent the homophily effects of drug usage, involvement in sports and the smoking behavior and the last term is an interaction between sport and Alcohol usage.

*Method 1:* We consider drug usage to be risky behavior, and assign the privacy risk of a dyad to depend on drug usage. We use two values for the probability of perturbing an edge: $\pi_1 = 38\%$ (i.e., privacy risk $\epsilon = 0.5$) if both nodes $i$ and $j$ use drugs, and $\pi_2 = 12\%$ (privacy risk $\epsilon = 2$), otherwise. Like with Enron data, the overall privacy risk for any dyad is measured by larger $\epsilon = 2$, but the privacy risk for dyads between nodes that use drugs is 0.5 (lower privacy risk) and hence their protection is higher. We compare this synthetic network release strategy with two others where all dyads are released either with $\pi = \pi_1$ or with $\pi = \pi_2$. Figure 5a shows the utility is lowest when $\pi = 38\%$ for all dyads which corresponds to the lowest disclosure risk while if we increase the risks for dyads between nodes that do not use drugs, we get an improved utility. If we use $\pi = 12\%$ for all dyads, we get much higher utility but at the expense of reduced privacy. The variation in the KL divergence from the missing data method is larger, by design, when $\pi = 38\%$, since this adds and removes a substantial number of edges and the missing data method reflects this additional uncertainly in the model estimates. This analysis also shows that assigning asymmetric privacy risk, even for more sparse networks, offers a better risk-utility trade-off.

*Method 2:* The privacy risk for every dyad is assumed to be the same ($1 - p_{ij} = 1 - q_{ij} = \pi$). Figures 5b, 5c and 5d show the KL divergence, MSE and bias results, and Figure 4 shows 8 synthetic networks generated using $\pi = 2\%$ while Table 8.2 shows the

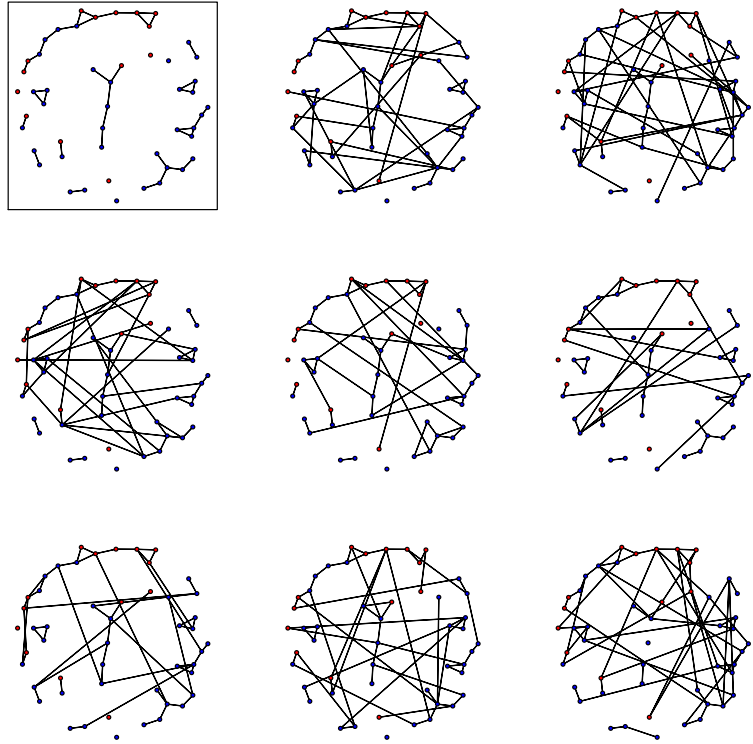

**Fig. 4.** 8 Synthetic copies of the Teenage Friendship Network using $\pi = 2\%$ and the original network (top left corner) are plotted in the same coordinate system for ease of comparison. Notice the addition of fake ties and removal of existing ties.

mean parameter estimates with their bias and the MSE. The results are very similar to those obtained in the Enron and Lazega case studies such that the missing data likelihood method outperforms the naive method by providing synthetic estimates closer to the original estimates (e.g., see Figure 5b with KL divergence especially for stronger privacy protection). The MSE of the missing data estimates in some cases is larger than the MSE of the naive estimates (e.g., homophily effect for sports). This is expected as the missing data estimates take into account the additional uncertain in the privacy mechanism. A few notable differences between the Teenage dataset, and for example, the Lazega dataset is the computed MSE and bias in the structural parameters, *GWESP*, *degree popularity* and the number of edges. In the Teenage data, the improvement in bias and the MSE of the structural parameters by using the missing data likelihood

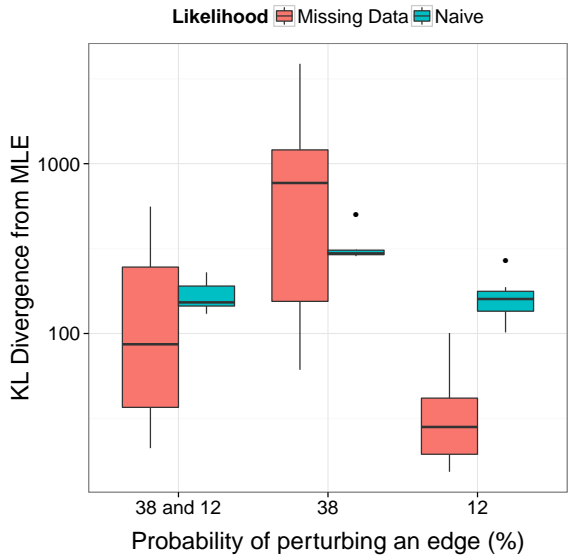

(a) Box plots of utility when privacy risks depend on the node attributes.

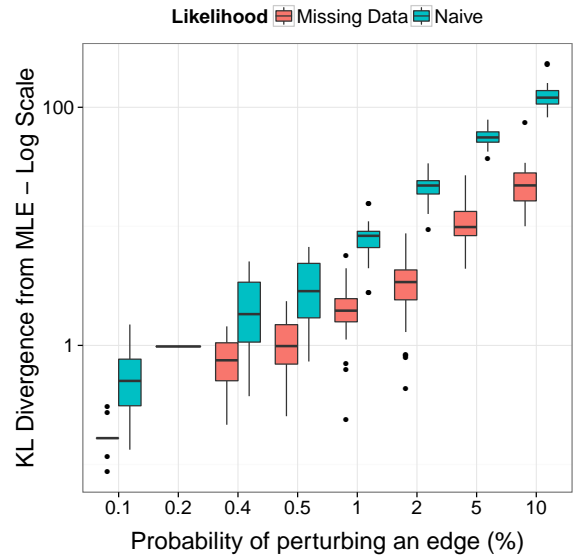

(b) KL divergence

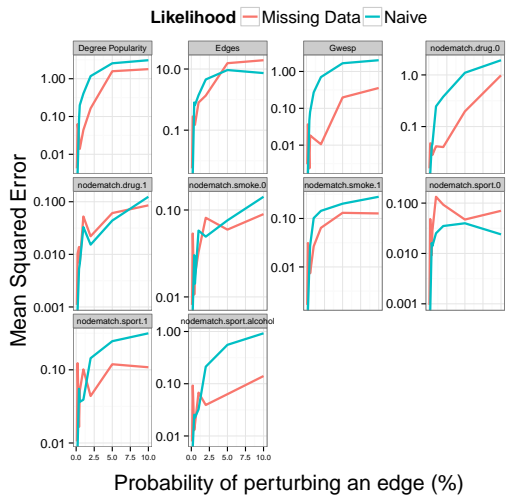

(c) Mean Squared Error

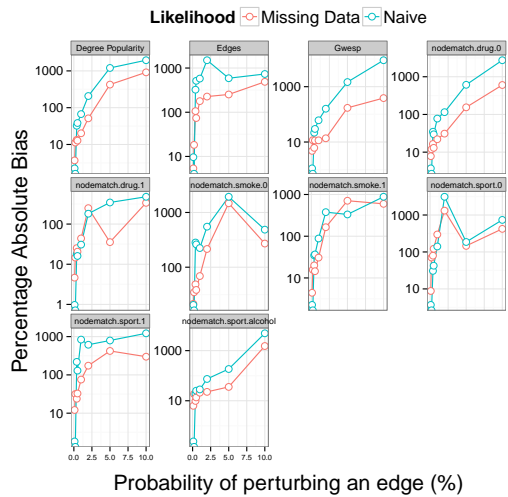

(d) Percentage Absolute Bias of parameters

**Fig. 5.** Teenage data: (a) and (b) Box plots of KL divergence of models obtained using the synthetic networks. The $x$-axis represents the probability of perturbing an edge in percentage (higher corresponds to more privacy risk). The $y$-axis shows the KL divergence in log scale (lower value corresponds to higher utility). The red color (lower value of KL) represents the KL divergence between the missing data estimates the MLE obtained from the original data. The cyan line (higher value of KL) represents the KL divergence between the naive estimates and the MLE. (c) MSE; (d) Absolute Bias of parameter estimates.

**Table 4.** Table showing the parameter estimates based on the original data (MLE) and the synthetic networks (Missing and Naive) for the Teenage Friendship Data. The probability of perturbing an edge is $2\%$.

| Parameter | MLE | Missing Data | | | Naive Data | | |
|---|---|---|---|---|---|---|---|
| | | estimate | mse | bias | estimate | mse | bias |
| Degree Popularity | -1.9 | -1.5 | 0.36 | 0.4 | -0.83 | 1.2 | 1.07 |
| Edges | 2.09 | 1.09 | 4.35 | -0.99 | 0.06 | 5.38 | -2.02 |
| gwesp.fixed.0 | 1.5 | 1.35 | 0.08 | -0.15 | 0.6 | 0.82 | -0.9 |
| nodematch.drug.0 | 1.57 | 1.74 | 0.28 | 0.17 | 0.97 | 0.47 | -0.59 |
| nodematch.drug.1 | 0.81 | 0.86 | 0.2 | 0.05 | 0.73 | 0.06 | -0.08 |
| nodematch.smoke.0 | -0.4 | -0.57 | 0.22 | -0.16 | -0.11 | 0.19 | 0.3 |
| nodematch.smoke.1 | 0.95 | 0.84 | 0.26 | -0.11 | 0.56 | 0.26 | -0.4 |
| nodematch.sport.0 | 0.53 | 0.36 | 0.49 | -0.17 | 0.47 | 0.2 | -0.06 |
| nodematch.sport.1 | -0.73 | -1.11 | 0.52 | -0.38 | -0.42 | 0.29 | 0.3 |
| nodematch.sport.alcohol | 1.31 | 1.42 | 0.17 | 0.1 | 0.83 | 0.28 | -0.48 |

method is not as high as the improvement obtained in the Lazega data. This is due to the fact that the Teenage data are much sparser when compared to the Lazega dataset.

*Acknowledgments.*

This work was supported in part by NSF grants BCS-0941553 and SES-1534433 to the Department of Statistics, Pennsylvania State University.